\documentclass[a4paper]{article}

\usepackage{INTERSPEECH2021}
\usepackage[ruled]{algorithm2e}

\title{Binary Neural Network for Speaker Verification}
\name{Tinglong Zhu$^1$, Xiaoyi Qin$^{1,2}$, Ming Li\thanks{Corresponding author: Ming Li}$^{1,2}$}
\address{
	$^1$Data Science Research Center, Duke Kunshan University, Kunshan, China\\
	$^2$School of Computer Science, Wuhan University, Wuhan, China}
\email{\{tinglong.zhu, xiaoyi.qin, ming.li369\}@dukekunshan.edu.cn}

\begin{document}

\maketitle
\begin{abstract}
Although deep neural networks are successful for many tasks in the speech domain, the high computational and memory costs of deep neural networks make it difficult to directly deploy high-performance Neural Network systems on low-resource embedded devices. There are several mechanisms to reduce the size of the neural networks i.e. parameter pruning, parameter quantization, etc. This paper focuses on how to apply binary neural networks to the task of speaker verification. The proposed binarization of training parameters can largely maintain the performance while significantly reducing storage space requirements and computational costs. Experiment results show that, after binarizing the Convolutional Neural Network, the ResNet34-based network achieves an EER of around 5\% on the \textbf{Voxceleb1} testing dataset and even outperforms the traditional real number network on the text-dependent dataset: \textbf{Xiaole} while having a 32x memory saving.
\end{abstract}

\section{Introduction}

In the past few years, the performance of speaker verification systems has been improved tremendously \cite{qian2019binary} due to the rapid development of neural networks, including Residual Neural Networks (ResNet)\cite{he2016deep}, Time Delay Neural Network (TDNN) \cite{peddinti2015time}, long-short term memory recurrent neural network (LSTM-RNNs), etc.

These neural network based systems outperform the traditional \textbf{i-vector} systems on several benchmark databases in terms of equal error rate (EER)and minimum detection cost function(minDCF)\cite{snyder2017deep}.
While the deeper convolutional neural networks (CNN) greatly improves the performance of systems on the speaker verification task, it requires considerable computational power and memory for both training process and inference stage, which results in that those systems perform well on GPU-based machines, but being infeasible to be deployed on low-resouce embedded devices\cite{rastegari2016xnor}. Thus, some recent studies are focusing on how to speed up the inference of deep neural networks and reduce the size of the model while keeping the performance degradation in a small and acceptable range\cite{qian2019binary}.

Prior studies have shown that there exists large redundancy in the deep neural network structure\cite{izui1990analysis,cheng2015exploration}, and the redundant computation will lead to a waste of computational resources. Thus it is possible for one to maintain the performance of a deep-network based system while reduce a large portion of its model's parameters\cite{qin2020binary}. 
Up till now, the approaches aimed at compressing the neural networks can be mainly classified into several categories\cite{qin2020binary}: parameter pruning\cite{cheng2015exploration, han2015deep,han2015learning}, parameter quantizing\cite{gong2014compressing,  wu2016quantized,vanhoucke2011improving}, low-rank parameter factorization\cite{denton2014exploiting, lebedev2014speeding}, transferred/compact convolutional filters\cite{howard2017mobilenets, sandler2018mobilenetv2}, and knowledge distillation\cite{hinton2015distilling, xu2017training}. For parameter pruning and quantizing, they mainly focus on parameter pruning and cutting down the parameter space. 
In prior researches, researchers have tried applying neural network compressing mechanisms to speaker verification systems. Those studies mainly focused on applying knowledge distillation mechanisms to their systems\cite{wang2019knowledge,mingote2020knowledge}. In this paper, we apply parameter binarization mechanism, which is a derivative of parameter quantization to compress the neural network in the speaker verification system.

Binarized parameters can have two possible values -1(0) or +1. As all the values can be represented within a single bits, the memory cost for inferencing the neural network model will be largely reduced. In addition, all the element-wise multiplication in the matrix computation can be simplified as addition/subtraction, which will also reduce the computational cost at the inference stage. Previous works in image processing community like BNN\cite{hubara2016binarized} and XNOR-Net\cite{rastegari2016xnor} have shown that the binarized networks can keep an acceptable accuracy while reducing ~32$\times$ memory costs at the inference stage. Furhtermore, if the binarizing certain layers would greatly degrade the performance of the networks we can conclude that these layers are the critical paths, in other words, these layers are not redundant\cite{qin2020binary}. Thus the study of binarizing the neural networks can also help the development of the full-precision neural networks. 

In this paper, we applied the Binary-Weight-Networks (BWN)\cite{rastegari2016xnor} quantization which uses +1/-1 as its parameter values.

The main contribution points are summarized as follows:
\begin{itemize}
\item We binarize the frontend network of the original speaker verification system. We only binarize the parameters of the frontend convolutional layers while keep the input and the gradient update using full precision number calculation.
\item We perform experiments on both text-dependent speaker verfication datasets \emph{Xiaole}\cite{jia20212020} and text-independent speaker verification \emph{Voxceleb1}\cite{Nagrani17}.
\item Based on the experimental results, we found that using ReLU function as the activation function performs better comparing to applying PReLU\cite{he2015delving} mentioned in prior studies\cite{kim2020improving} of binary neural networks.
\item We compute the total size of the binarized parameters and find that we can compress the space to 1/32 of the original size if the corresponding underlying code for binarization is well implemented.
\end{itemize}

\section{Binary Weight Neural Networks}

In this section, we will introduce the binarization function we used, how we apply it to the parameter gradients calculation, how we backpropagate the loss and the way we update the model parameters.  

\begin{figure*}[t]
	\centering
	\includegraphics[width=\linewidth]{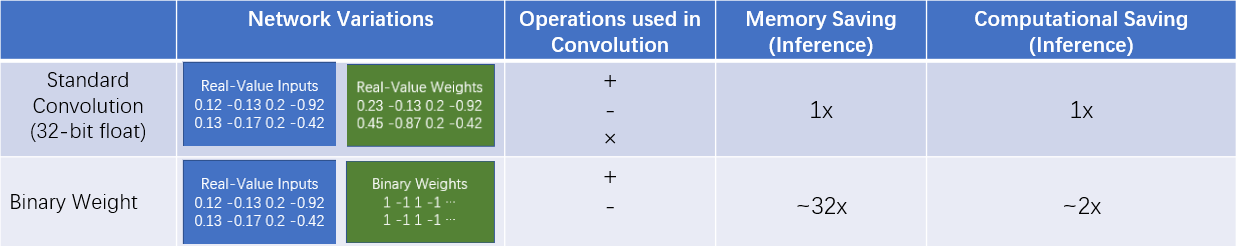}
	\caption{Binary Weight Networks}
	\label{fig:bwn}
\end{figure*}

\subsection{Binarization Function}
In order to reduce the model size, we constrain model parameters to -1/+1,as those two values are competitive under a hardware perspective\cite{hubara2016binarized}. There are two main binarization functions in the prior studies\cite{qin2020binary,hubara2016binarized}:
\begin{equation}
x^b= {\rm Sign}(x)=
	\left\{
		\begin{array}{lcl}
			&+1 \quad& {\rm if} x \geq 0\\
			&-1  \quad& {\rm otherwise} \\
		\end{array}
	\right.
	\label{eq1}
\end{equation}
\begin{equation}
x^b=
	\left\{
		\begin{array}{lcl}
			&+1 &{\rm with\  probability } \  p =\sigma(x)\\
			&-1  &{\rm with\  probability }\ 1-p \\
		\end{array}
	\right.
\label{eq2}
\end{equation}
where $\sigma$ in (\ref{eq2}) is the \emph{"hard sigmoid"} function:
\begin{equation}
	\sigma(x)=  {\rm clip}(\frac{x+1}{2},0,1)={\rm max}(0,{\rm min}(1, \frac{x+1}{2}))
	\label{eq3}
\end{equation}
In this paper, like \cite{hubara2016binarized} we apply equation (\ref{eq1}) as our binarization function as it is the optimal solution of the binary operation mentioned in the following section. Meanwhile, (\ref{eq1}) is also easier to be implemented and costs less computational resources in both training and inferencing.
\subsection{Binarization Process}
First, we denote the input tensor of a CNN as \textbf{I}. (\textbf{I} $ \in \mathbb{R}^{c,w,h}$), where $(c,w,h)$ represents channels, width and height respectively.\textbf{W} is the set of the weight filter of a single layer of CNN. (\textbf{W} $ \in \mathbb{R}^{c,w_0,h_0}$, where $w_0\leq w$,$h_0\leq h$).
\subsection{Convolutional Layer Binarization}\label{convlayer}
A binary weight network\cite{rastegari2016xnor} uses a binary filter \textbf{B} $\in \{+1,-1\}^{c\times w_0 \times h_0} $ and a scaling factor $a \in \mathbb{R}^+$ such that we can simulate the real number filters by using $\textbf{W} \approx a\textbf{B}$ Thus, a convolutional operation can be approximated by:
\begin{equation}
	\textbf{I} * \textbf{W} \approx (\textbf{I} \oplus \textbf{B} )a
	\label{eq4}
\end{equation}
where $\oplus$ represents a convolution computation without any multiplication. As all the weight values are either +1/-1, then all the operations can be viewed as additions and subtractions. 

In order to keep the performance of the binary neural networks as much close to the real number neural networks as possible, it is necessary for the binary convolutional operation as much close to the full precision convolutional operation as possible. Thus the optimization problem would be transformed into\cite{rastegari2016xnor}:

\begin{equation}
	J(\textbf{B},a)=\left\| \textbf{W}-a\textbf{B}\right\|_2 \qquad
	a^*,\textbf{B}^*= \mathop{\arg\min}\limits_{a,\textbf{B}}J(\textbf{B},a)
	\label{eq5}
\end{equation}
By expanding the (\ref{eq5}): 
\begin{equation}
	J(\textbf{B},a) = a^2\textbf{B}^T\textbf{B}- 2a \textbf{W}^T\textbf{B}  + \textbf{W}^T\textbf{W}
\end{equation}
as $\textbf{B}_i \in \{-1,+1\}^{d}, {\rm where } \  d\ {\rm is}\  c\times w_0 \times h_0$, thus $\textbf{B}^T\textbf{B}=n$ is a constant. At the same time, as $\textbf{W}$ is the weight of a single convolutional layer, it is also known, $s.t.$ $\textbf{W}^T \textbf{W} =k$ can be treated as a constant, too. Then (\ref{eq5}) can be rewritten as $J(\textbf{B},a) =a^2n-2a \textbf{W} ^T \textbf{B} + k$. Thus, the optimization problem for $\textbf{B}$  can be transformed to 
\begin{equation}
	\textbf{B}^* = \mathop{\arg\max}\limits_{\textbf{B}}\{\textbf{W}^T\textbf{B}\}\quad s.t. \textbf{B} \in \{+1,-1\}^{c\times w_0\times h_0}
	\label{eq6}
\end{equation}
Therefore, the solution\cite{rastegari2016xnor} for (\ref{eq6}) is:
\begin{equation}
	\textbf{B}_i = {\rm Sign} (\textbf{B}_i)
	\label{eq7}
\end{equation}
Based on (\ref{eq7}) we can get 
\begin{equation}
	a^*=\frac{\textbf{W}^T {\rm Sign}(\textbf{W})}{n} = \frac{1}{n} \left\| \textbf{W} \right\|_{l1}
\end{equation}
In other words, the optimal solution for $a$ is the average value of the weight values.

\begin{figure}[t]
	\centering
	\includegraphics[width=0.8\linewidth]{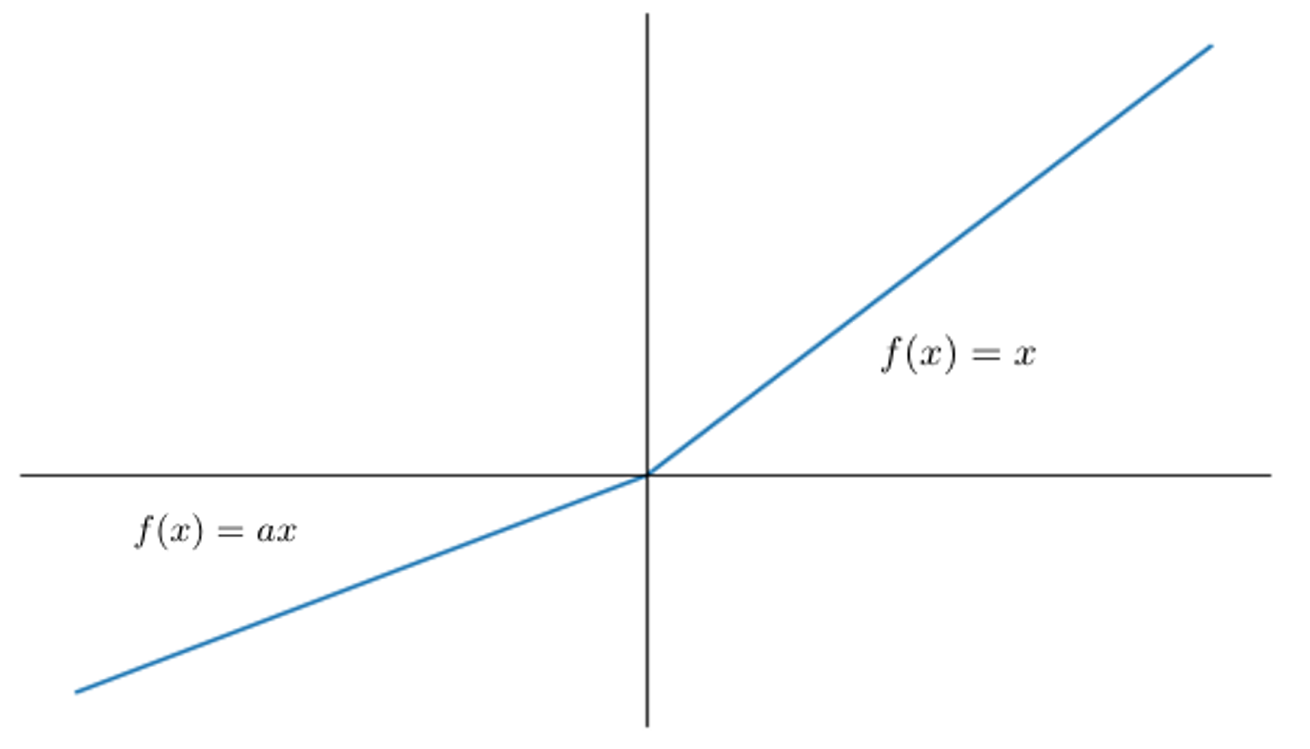}
	\caption{PReLU Activation}
	\label{fig:prelu}
\end{figure}

\subsection{Activation Function for Binary Weight Networks}
In the traditional ResNets, Rectified Linear Unit (ReLU) is applied as the activation function. In the prior study of the Binary Neural Networks like\cite{kim2020improving}, they applied PReLU as shown in \cite{he2015delving} Fig. \ref{fig:prelu} as the activation function. But based on our experiments, the PReLU activation has great negative effect on our binarized systems as shown in Table (\ref{tab:voxres}). Thus, we choose to apply ReLU as our activation function.

%

\subsection{Training the Binary Weight Neural Networks}

In this paper, we mainly use the method from \cite{rastegari2016xnor} train the Binary Weight Neural Networks. But there exists tiny difference. The steps of training a CNN can be splitted into 3 steps: forward pass, backward pass and parameters update. The whole training process is shown in Algorithm \ref{alg:bnn}and we will introduce the training steps in detail within the following sections.

\subsubsection{Forward}
In the initial Binary-Weights-Networks\cite{rastegari2016xnor}, the weights are binarized in the forward process. But based on our experiments, not binarizing the weights in this process will bring better performance. As not binarizing the weights in this process would keep the information of the full precision neural networks to the greatest extent. 

\subsubsection{Backward and Parameters Update}
We use the same approach for calculating gradient for the Sign function \cite{hubara2016binarized}. That is, when the binary quantization is
$x^b = {\rm Sign} (x)$. Then the gradient $\frac{\partial C}{\partial x}$ is 
\begin{equation}
	g_x=g_{x^b}1_{|r|\leq 1}
\end{equation}
The gradient in the backward process would be calculated as 
\begin{equation}
	\frac{\partial C}{\partial W_i} = \frac{\partial C}{\partial\hat{W}}\left(\frac{1}{n}+ \frac{\partial{\rm Sign}}{\partial \widetilde{W_i}}\alpha\right)
\end{equation}
It is necessary to use full precision real number for updating the parameters\cite{rastegari2016xnor} as during gradient descending process, the changes for the parameters are tiny, if we do not apply the high precision numebr updating stategy, then the binarization step would ignore the changes, which may lead the network remain unupdated. 
\begin{algorithm}[h]
	\caption{Binary Weight CNN Training}
	\label{alg:bnn}
	\LinesNumbered 
	\KwIn{A mini batch of inputs and targets (\textbf{I,Y}), cost function $C$ (\textbf{Y,$\hat{\textbf{Y}}$}), current weight $W^t$ and current learning rate $\eta^t$}
	\KwOut{weight $W^{t+1}$ and learning rate $\eta^{t+1}$ after update}
	
	$\hat{Y} = \textbf{Forward} (\textbf{I},W^t)$ // Here we do not use binary forward as it was in \cite{rastegari2016xnor}. We just use standard forward propagation.\;
	Binarize weight filters, $L$ layers in total\; 
	\For{ $l=1 $ to $L$ }{
		\For{ $k^{th}$ filter in $ l^{th} $layer}{
			$A_{lk}=\frac{1}{n}\left\|W_{lk}^t \right\|$\;
			$B_{lk}={\rm Sign} (W_{lk}^t)$\;
			$\widetilde{W_{lk}^t}=A_{lk}B_{lk}$\;
		}
	}
	$\frac{\partial C}{\partial \widetilde{W}} = \textbf{BinaryBackward} (\frac{\partial C}{\partial \hat{Y}},\widetilde{W})$ // standard backward propagation except that gradients are computed using $\widetilde{W}$ other than $W^t$\cite{rastegari2016xnor}\;
	$W^{t+1}= \textbf{UpdateParameters}(W^t,\frac{\partial C}{\widetilde{W}},\eta^t)$\;
	$\eta^{t+1}=\textbf{UpdateLearningrate}(\eta^t,t)$\;
	
\end{algorithm}
\subsubsection{Binary Training Algorithm}
For the training process described in Algorithm \ref{alg:bnn}, we apply the similar procedure as that in\cite{rastegari2016xnor}. First, we use the standard forward propagation (full precision real number) to calculate $\hat{Y}$: predicted speaker's label. Then we binarize the weight filters at each layer by computing $B$ (sign of weight parameters) and $A$ (average value of weight parameters). After that, we do the backward propagation (using the binarized weight filters $\widetilde{W}$ to calculate the gradients). For the last step, we apply the standard update rules for parameters and learning rate. In this work, we use \textbf{SGD} optimizer with \textbf{momentum} to update parameters.

\section{Experiments}

\subsection{Experiments Setup}
\subsubsection{Datasets.} We use \textbf{Voxceleb1} dev\cite{Nagrani17} as text-independent speaker verification training dataset, which contains over 1200 speakers and over 140,000 utterances. For text-independent speaker verification testing, we use \textbf{Voxceleb1} test dataset, which contains 40 speakes and over 4000 utterances. For the text-dependent speaker verification, we use part of \textbf{Xiaole}\cite{jia20212020} dataset. \textbf{Xiaole} dataset has 550 speakers in total. In this paper, we only use close-talk utterances in \textbf{Xiaole} for training and testing. The training set we used contains around 300 speakers and over 34,000 close-talk utterances from the training dataset of \textbf{Xiaole}, while the dataset we used for testing contains 97 speakers and over 11,000 close-talk utterances from the testing dataset of \textbf{Xiaole}.

\subsubsection{Implementation Details.} We use Pytorch to implement all our experiments. We've adopted binarized ResNet34, and TDNN, as the frontend embedding module, and neural network based Classifier with CrossEntropy or GE2E loss for training. For back-end scoring, we tried both Probabilistic Linear Discriminant Analysis (PLDA) and computing the cosine similarity between two embeddings for verfication. For both ResNet and TDNN, we adopt the SGD optimizer with a momentum factor of 0.95, set the starting learning rate to be 0.01, and the learning rate will decay 90\% every 10 epochs. For experiments with GE2E loss, we set the batch size to 40, with 5 speaker, who each has 8 utterances in the batch. For the ResNet34, we did not apply dropout strategy and we set the embedding size to be 128.
\begin{table*}[htpb]
	\center
	
	\caption{Text-independent speaker verification experiments on Voxceleb1}
	\label{tab:voxres}
	\begin{tabular}{lllll}
		\toprule
		\textbf{Model}    & \textbf{Loss}  & \textbf{Back-end}       & \textbf{Activation} & \textbf{EER/minDCF}    \\ 
		
		\midrule
		ResNet34 (float32) & CrossEntropy & Cosine Similarity        & ReLU       & 3.31\%/-      \\
		ResNet34 (float32) & CrossEntropy & PLDA        & ReLU       & 3.26\%/-      \\
		ResNet34 (binary)      & GE2E         & Cosine Similarity        & ReLU       & 9.724\%/0.642  \\
		ResNet34 (binary)      & GE2E         & PLDA        & ReLU       & 7.89\%/0.674  \\
		ResNet34   (binary)    & CrossEntropy & Cosine Similarity        & ReLU       & 9.968\%/0.794 \\
		ResNet34   (binary)    & CrossEntropy & Cosine Similarity        & PReLU       & 24.973\%/0.90 \\
		ResNet34    (binary)   & CrossEntropy & PLDA     & ReLU       & \textbf{5.355\%/0.520} \\
		BIG\_ETDNN    (binary) & CrossEntropy & Cosine Similarity        & ReLU       & 9.841\%/0.530 \\ 
		\bottomrule
	\end{tabular}
	
\end{table*}
\begin{table*}[htpb]
	\center
	
	\caption{Text-dependent speaker verification experiments on Xiaole}
	\label{tab:xiaole}
	\begin{tabular}{lllll}
		\toprule
		\textbf{model}    & \textbf{Loss}  & \textbf{Back-end}       & \textbf{Activation} & \textbf{EER/minDCF}    \\ 
		
		\midrule
		ResNet34 (float32)  & CrossEntropy & Cosine Similarity        & ReLU       & 4.464\%/0.406     \\
		ResNet34 (binary) & CrossEntropy & Cosine Similarity        & ReLU       & 3.338\%/0.456 \\
		ResNet34 (binary) & CrossEntropy & PLDA     & ReLU       & \textbf{3.098\%/0.321} \\ 
		\bottomrule
	\end{tabular}
	
\end{table*}


\subsection{Text-independent Speaker Verification}

The baseline in Table \ref{tab:voxres}, is based on the end-to-end speaker verification system proposed in \cite{cai2018exploring}. Based on the equal error rate (EER) and minimum detection cost function (minDCF), we could observe that: if we only binarize the original CNN based speaker verification system, the performance of the system would be largely degraded. But with the help of the back-end PLDA scoring, the binarized system can achieve similar performance comparing to the full precision real number systems. Obviously, PLDA have tremendous improvement on the binarized system, while it has little affect on the full precision real number based system. One reason may be that: binarized systems have much smaller model-space than that of the full-precision number based systems, thus it has weaker performance in recognizing the difference between different speakers. As the PLDA scoring helps to minimize the distance between the embeddings of the same speaker while maximizing the distance between the embeddings of different speakers, PLDA improves the capability of recognizing different speakers of the binarized system. In addition, the end-to-end i.e. GE2E loss would also help to improve the binarized system, and the system trained on GE2E loss can also take advantage of the PLDA.
\begin{table}[]
	
	\caption{Parameters and Size of Binary Weight Networks}
	\label{tab:param}
	\centering
	\begin{tabular}{lll}
		\toprule
		\textbf{Model}              & \textbf{Params} & \textbf{Precision}         \\ 
		\midrule
		ResNet34   & 21.6M      &float32     \\
		ResNet34   &  675K &binary \\
		BIG\_ETDNN  & 18.7M &float32          \\
		BIG\_ETDNN &  584K &binary \\ 
		\bottomrule
	\end{tabular}
\end{table}
\subsection{Text-dependent Speaker Verification}
We apply the same system with text-independent speaker verification\cite{cai2018exploring} to create the baseline for the text-dependent speaker verification task. From Table \ref{tab:xiaole}, it is clear that the performance of the binarized CNN system is over 20\% 
better than that of the real number based systems. With the help of the PLDA scoring, the system performs even better. The better performance of the binarized system on text-dependent speaker verification task may due to the less requirements of the model space of the text-dependent verification task.
\subsection{Size of Binarized Model}
Though the real number based CNN systems are best known for its high performance, the size of the CNN based systems are so large that they cannot easily be applied to the low-resource embedded devices. In Table \ref{tab:param}, we find that, with appropriate implementation, the space needed for the binarized model is about 1/32 less than that of the real number based systems. Which makes the binarized systems possible to be applied to the embedding devices while keeping the similar performance comparing to the real number systems.
\begin{figure}[t]
	\centering
	\includegraphics[width=\linewidth]{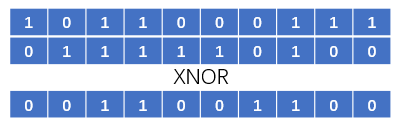}
	\caption{XNOR Bit Wise Computation}
	\label{fig:xnor}
\end{figure}
\subsection{Computational Cost of Binarized Model}
We did binarize the weight filters of the Convolutional Neural Network layer, but as we did not binarize the input as \cite{rastegari2016xnor} did. As a result, our system still need to apply floating points calculation other than bit-wise computation in \cite{rastegari2016xnor} as shown by Fig. \ref{fig:xnor}. But as all the weights are either +1/-1 as mentioned in section \ref{convlayer}, the computation can be simplified from multiplications to additions and subtractions. As the costs for the multiplication is higher than additions/subtractions for some low-resource embedded devices, thus our system will also reduce the computational costs at the inference stage.
\section{Conclusion}
In this paper, we show that Binary-Weight-Networks can be applied to speaker verification systems while maintaining its performance to the large extent. By binarizing the frontend CNN model, we dramatically reduce the space requirement of some high performing speaker verification systems. But the problem brought by large computational cost still remains, as the inputs of our binarized systems are still real numbers. So that in the inference process, the binarized parameters will be cast to the full precision float numbers for calculation. Even though the calculation is still floating calculation, as mentioned above, all the calculation can be cast to the addition and substraction operation. Thus with proper implementation, theoretically, the computational costs at inference stage can be reduced at least 50\% as shown in Fig. \ref{fig:bwn}. For the future studies, we could try to apply Xnor-Net \cite{rastegari2016xnor} to the speaker verification system to further reduce the computational cost.
\bibliographystyle{IEEEtran}
\bibliography{mybib}

\begin{thebibliography}{10}
\providecommand{\url}[1]{#1}
\csname url@samestyle\endcsname
\providecommand{\newblock}{\relax}
\providecommand{\bibinfo}[2]{#2}
\providecommand{\BIBentrySTDinterwordspacing}{\spaceskip=0pt\relax}
\providecommand{\BIBentryALTinterwordstretchfactor}{4}
\providecommand{\BIBentryALTinterwordspacing}{\spaceskip=\fontdimen2\font plus
\BIBentryALTinterwordstretchfactor\fontdimen3\font minus
  \fontdimen4\font\relax}
\providecommand{\BIBforeignlanguage}[2]{{%
\expandafter\ifx\csname l@#1\endcsname\relax
\typeout{** WARNING: IEEEtran.bst: No hyphenation pattern has been}%
\typeout{** loaded for the language `#1'. Using the pattern for}%
\typeout{** the default language instead.}%
\else
\language=\csname l@#1\endcsname
\fi
#2}}
\providecommand{\BIBdecl}{\relax}
\BIBdecl

\bibitem{qian2019binary}
Q.~Yanmin and X.~Xiang, ``Binary neural networks for speech recognition,''
  \emph{Frontiers of Information Technology \& Electronic Engineering},
  vol.~20, no.~5, pp. 701--715, 2019.

\bibitem{he2016deep}
K.~He, X.~Zhang, S.~Ren, and J.~Sun, ``Deep residual learning for image
  recognition,'' in \emph{Proceedings of the IEEE conference on computer vision
  and pattern recognition}, 2016, pp. 770--778.

\bibitem{peddinti2015time}
V.~Peddinti, D.~Povey, and S.~Khudanpur, ``A time delay neural network
  architecture for efficient modeling of long temporal contexts,'' in
  \emph{Sixteenth Annual Conference of the International Speech Communication
  Association}, 2015.

\bibitem{snyder2017deep}
D.~Snyder, D.~Garcia-Romero, D.~Povey, and S.~Khudanpur, ``Deep neural network
  embeddings for text-independent speaker verification.'' in
  \emph{Interspeech}, 2017, pp. 999--1003.

\bibitem{rastegari2016xnor}
M.~Rastegari, V.~Ordonez, J.~Redmon, and A.~Farhadi, ``Xnor-net: Imagenet
  classification using binary convolutional neural networks,'' in
  \emph{European conference on computer vision}.\hskip 1em plus 0.5em minus
  0.4em\relax Springer, 2016, pp. 525--542.

\bibitem{izui1990analysis}
Y.~Izui and A.~Pentland, ``Analysis of neural networks with redundancy,''
  \emph{Neural Computation}, vol.~2, no.~2, pp. 226--238, 1990.

\bibitem{cheng2015exploration}
Y.~Cheng, F.~X. Yu, R.~S. Feris, S.~Kumar, A.~Choudhary, and S.-F. Chang, ``An
  exploration of parameter redundancy in deep networks with circulant
  projections,'' in \emph{Proceedings of the IEEE international conference on
  computer vision}, 2015, pp. 2857--2865.

\bibitem{qin2020binary}
H.~Qin, R.~Gong, X.~Liu, X.~Bai, J.~Song, and N.~Sebe, ``Binary neural
  networks: A survey,'' \emph{Pattern Recognition}, vol. 105, p. 107281, 2020.

\bibitem{han2015deep}
S.~Han, H.~Mao, and J.~W. Dally, ``Deep compression: Compressing deep neural
  network with pruning, trained quantization and huffman coding,''
  \emph{international conference on learning representations}, 2015.

\bibitem{han2015learning}
S.~Han, J.~Pool, J.~Tran, and W.~Dally, ``Learning both weights and connections
  for efficient neural network,'' in \emph{Advances in Neural Information
  Processing Systems}, C.~Cortes, N.~Lawrence, D.~Lee, M.~Sugiyama, and
  R.~Garnett, Eds., vol.~28.\hskip 1em plus 0.5em minus 0.4em\relax Curran
  Associates, Inc., 2015.

\bibitem{gong2014compressing}
Y.~Gong, L.~Liu, M.~Yang, and D.~L. Bourdev, ``Compressing deep convolutional
  networks using vector quantization,'' \emph{Computing Research Repository},
  2014.

\bibitem{wu2016quantized}
J.~Wu, C.~Leng, Y.~Wang, Q.~Hu, and J.~Cheng, ``Quantized convolutional neural
  networks for mobile devices,'' in \emph{Proceedings of the IEEE Conference on
  Computer Vision and Pattern Recognition}, 2016, pp. 4820--4828.

\bibitem{vanhoucke2011improving}
V.~Vanhoucke, A.~Senior, and M.~Z. Mao, ``Improving the speed of neural
  networks on cpus,'' 2011.

\bibitem{denton2014exploiting}
E.~Denton, W.~Zaremba, J.~Bruna, Y.~LeCun, and R.~Fergus, ``Exploiting linear
  structure within convolutional networks for efficient evaluation,''
  \emph{arXiv preprint arXiv:1404.0736}, 2014.

\bibitem{lebedev2014speeding}
V.~Lebedev, Y.~Ganin, M.~Rakhuba, I.~Oseledets, and V.~Lempitsky, ``Speeding-up
  convolutional neural networks using fine-tuned cp-decomposition,''
  \emph{arXiv preprint arXiv:1412.6553}, 2014.

\bibitem{howard2017mobilenets}
A.~G. Howard, M.~Zhu, B.~Chen, D.~Kalenichenko, W.~Wang, T.~Weyand,
  M.~Andreetto, and H.~Adam, ``Mobilenets: Efficient convolutional neural
  networks for mobile vision applications,'' \emph{arXiv preprint
  arXiv:1704.04861}, 2017.

\bibitem{sandler2018mobilenetv2}
M.~Sandler, A.~Howard, M.~Zhu, A.~Zhmoginov, and L.-C. Chen, ``Mobilenetv2:
  Inverted residuals and linear bottlenecks,'' in \emph{Proceedings of the IEEE
  conference on computer vision and pattern recognition}, 2018, pp. 4510--4520.

\bibitem{hinton2015distilling}
E.~G. Hinton, O.~Vinyals, and J.~Dean, ``Distilling the knowledge in a neural
  network,'' \emph{Computing Research Repository}, 2015.

\bibitem{xu2017training}
Z.~Xu, Y.-C. Hsu, and J.~Huang, ``Training shallow and thin networks for
  acceleration via knowledge distillation with conditional adversarial
  networks,'' \emph{international conference on learning representations},
  2018.

\bibitem{wang2019knowledge}
S.~Wang, Y.~Yang, T.~Wang, Y.~Qian, and K.~Yu, ``Knowledge distillation for
  small foot-print deep speaker embedding,'' in \emph{ICASSP 2019-2019 IEEE
  International Conference on Acoustics, Speech and Signal Processing
  (ICASSP)}.\hskip 1em plus 0.5em minus 0.4em\relax IEEE, 2019, pp. 6021--6025.

\bibitem{mingote2020knowledge}
V.~Mingote, A.~Miguel, D.~Ribas, A.~Ortega, and E.~Lleida, ``Knowledge
  distillation and random erasing data augmentation for text-dependent speaker
  verification,'' in \emph{ICASSP 2020-2020 IEEE International Conference on
  Acoustics, Speech and Signal Processing (ICASSP)}.\hskip 1em plus 0.5em minus
  0.4em\relax IEEE, 2020, pp. 6824--6828.

\bibitem{hubara2016binarized}
M.~Courbariaux and Y.~Bengio, ``Binarynet: Training deep neural networks with
  weights and activations constrained to +1 or -1,'' \emph{Computing Research
  Repository}, 2016.

\bibitem{jia20212020}
Y.~Jia, X.~Wang, X.~Qin, Y.~Zhang, X.~Wang, J.~Wang, and M.~Li, ``The 2020
  personalized voice trigger challenge: Open database, evaluation metrics and
  the baseline systems,'' \emph{arXiv preprint arXiv:2101.01935}, 2021.

\bibitem{Nagrani17}
A.~Nagrani, J.~S. Chung, and A.~Zisserman, ``Voxceleb: a large-scale speaker
  identification dataset,'' in \emph{INTERSPEECH}, 2017.

\bibitem{he2015delving}
K.~He, X.~Zhang, S.~Ren, and J.~Sun, ``Delving deep into rectifiers: Surpassing
  human-level performance on imagenet classification,'' in \emph{Proceedings of
  the IEEE international conference on computer vision}, 2015, pp. 1026--1034.

\bibitem{kim2020improving}
H.~Kim, J.~Park, C.~Lee, and J.-J. Kim, ``Improving accuracy of binary neural
  networks using unbalanced activation distribution,'' \emph{arXiv preprint
  arXiv:2012.00938}, 2020.

\bibitem{cai2018exploring}
W.~Cai, J.~Chen, and M.~Li, ``Exploring the encoding layer and loss function in
  end-to-end speaker and language recognition system,'' \emph{Odyssey}, pp.
  74--81, 2018.

\end{thebibliography}


\end{document}